\documentclass[useAMS,usenatbib,usegraphicx]{mn2e}

\usepackage{amssymb,amsmath}

\newcommand{\aap}{A\&A}          
\newcommand{\apj}{ApJ}           
\newcommand{\apjs}{ApJS}         
\newcommand{\apjl}{ApJL}         
\newcommand{\mnras}{MNRAS}       
\newcommand{\pasp}{PASP}         
\newcommand{\nat}{Nature}

\newcommand{\jqsrt}{Journal of Quantitative Spectroscopy and Radiative Transfer}

\newcommand{\kms}{\hbox{km s$^{-1}$}}
\newcommand{\re}{\hbox{${\rm R}_{\rm e}$}} 

\newcommand{\msun}{\hbox{M$_{\odot}$}}
\newcommand{\mydef}{\mathrel{\overset{\makebox[0pt]{\mbox{\normalfont\tiny\sffamily def}}}{=}}}

\title[The IMF of a massive relic galaxy]{The initial mass function of a massive relic galaxy}
\author[Mart\'in-Navarro et al.]{Ignacio Mart\'in-Navarro$^{1,2}$\thanks{E-mail: \textit{I. Mart\'in-Navarro} imartin@iac.es}, Francesco La Barbera$^{3}$, Alexandre Vazdekis$^{1,2}$, 
\newauthor
Anna Ferr\'e-Mateu$^{4}$, Ignacio Trujillo$^{1,2}$ \& Michael A. Beasley$^{1,2}$\\ 
$^{1}$Instituto de Astrof\'isica de Canarias, E-38200 La Laguna, Tenerife, Spain\\
$^{2}$Departamento de Astrof\'isica, Universidad de La Laguna, E-38205 La Laguna, Tenerife, Spain\\
$^{3}$INAF - Osservatorio Astronomico di Capodimonte, Napoli, Italy\\
$^{4}$Subaru Telescope, 650 North A'ohoku Place, Hilo, HI 96720, USA}

\begin{document}

\date{ }

\pagerange{\pageref{firstpage}--\pageref{lastpage}} \pubyear{2014}

\maketitle
\label{firstpage}

\begin{abstract}

Massive relic galaxies formed the bulk of their stellar component before $z\sim2$ and have remained unaltered since then. Therefore, they represent a unique opportunity to study in great detail the \emph{frozen} stellar population properties of those galaxies that populated the primitive Universe. We have combined optical to near-infrared line-strength indices in order to infer, out to 1.5~\re, the IMF of the nearby relic massive galaxy NGC~1277. The IMF of this galaxy is bottom-heavy at all radii, with the fraction of low-mass stars being at least a factor of two larger than that found in the Milky Way. The excess of low-mass stars is present throughout the galaxy, while the velocity dispersion profile shows a strong decrease with radius. This behaviour suggests that local velocity dispersion is not the only driver of the observed IMF variations seen among nearby early-type galaxies. In addition, the excess of low-mass stars shown in NGC~1277 could reflect the effect on the IMF of dramatically different and intense star formation processes at $z\sim2$, compared to the less extreme conditions observed in the local Universe.
\end{abstract}


\section{Introduction}

A combination of major and minor dry mergers are the favoured explanation for the evolution of the scaling relations of massive galaxies since $z\sim2$ \citep[e.g.][]{naab,truji11,ferreras14,ruiz}. Being a stochastic process, a natural prediction of this evolutionary channel is the presence, in the local volume, of a number of massive galaxies which have experienced few if any significant merger events since they were formed \citep{Quilis}. These putative \textit{relic galaxies} present a unique opportunity to study in detail the properties of massive, $z\sim2$ galaxy analogues in the nearby Universe.

The Perseus cluster galaxy NGC~1277 fulfills all the characteristics to be considered a relic massive galaxy \citep{truji14}. The galaxy exhibits an unusually high  central velocity dispersion (above 400 \kms) and is physically compact, with an effective circularized radius of 1.2 kpc \citep{truji14}. The bulk of its stars are old (age $>$ 10 Gyr) and $\alpha$-enhanced ([Mg/Fe] $\sim$ 0.3 dex), which suggests a short ($\tau \sim 100$ Myr) and intense (SFR $\gtrsim 10^{3}$ \msun/yr) star formation burst at  $z \sim 2$, followed by subsequent quiescent evolution \citep{truji14}. In addition, NGC~1277 hosts the most massive (compared to the total galaxy mass) black hole found to date \citep[][but see \citealt{Emsellem}]{remco}.

\citet{truji14} studied the stellar population properties of NGC~1277 assuming  that its stellar initial mass function (IMF) follows that seen in the Milky Way \citep{kroupa,chabrier}. Although this approach has been extensively used in the literature, many studies \citep{cenarro,vandokkum,cappellari,ferreras,labarbera,Spiniello2013} have shown that the shape and the normalization of the IMF correlate with galaxy mass. Moreover, \citet{mn14} have found that the IMF slope  of some representative massive galaxies depends on galactocentric distance. Since the inferred properties of a stellar population strongly depend on the adopted IMF \citep{anna13}, it is important to investigate the stellar populations of massive systems such as NGC~1277 while relaxing the assumption of a universal IMF.

The importance of measuring the behaviour of the IMF as a function of radius in NGC~1277 is two-fold. On the one hand, it provides a route to understanding the physical conditions of massive galaxies when the majority of their stars were formed, with no influence from later accretion events. On the other hand, NGC~1277 shows an atypical combination of kinematical and stellar population properties \citep{remco,truji14}, which can be used to investigate the main driver behind these IMF variations.

By studying the radial variation of IMF-sensitive spectral indices in NGC~1277, we have been able to derive its IMF radial profile, which remains bottom-heavy  up to $\sim$ 1.5\re. We have simultaneously inferred, in a self-consistent manner (i.e., by adopting the best-fitting IMF value at each radius), its age, metallicity and [Mg/Fe] gradients. Here we find that the claims made by \citet{truji14} regarding NGC~1277 as being a relic object are also supported under the assumption of a steep IMF.

This paper is structured as follows: in Section 2 we present our data; Section 3 contains the analysis and the main results, which are finally discussed in Section 4. We adopt a standard cosmology: H$_0$= 70 km s$^{-1}$ Mpc$^{-1}$, $\Omega_m$= 0.3, and $\Omega_\Lambda$=0.7.

\section{Data and data reduction} \label{sec:data}

Our data consist of two sets of long-slit spectroscopic observations of NGC~1277 carried out at the 4.2m William Hershel Telescope (WHT) and at the 10.4m Gran Telescopio Canarias (GTC), located at the Spanish Observatorio del Roque de los Muchachos on La Palma. The WHT spectra were acquired with the Intermediate dispersion Spectrograph and Imaging System (ISIS), using the  R300B grating and a 1~arcsec slit, placed along the major axis of the galaxy \citep[see][for details]{truji14}. The total on-source integration time for these observations was 3 hours, with 1~arcsec seeing. The  spectra cover the wavelength range from $\sim$ 3700 to $\sim$ 6000 \AA, with 3.4 \AA \ spectral resolution (FWHM). In a second run of observations, we again targeted the NGC~1277 major axis with the GTC Optical System for Imaging and low-Intermediate-Resolution Integrated Spectroscopy (OSIRIS) in order to increase the wavelength coverage up to 10000\,\AA, allowing us to include near-infrared IMF-sensitive features such as Na~8190 \citep{Schiavon:00} and the Calcium triplet \citep{cat}. These data have a (wavelength-dependent) resolution ranging from $6$\,\AA \ at $\lambda \sim 5000$\,\AA \ to $8$\,\AA \ at $\lambda\sim 10000$\,\AA. The total exposure time was 1.5 hours on source with a typical 0.8 arcsec seeing. In both datasets, the pixel size was 0.25 arcsec. In this work, all measurements (i.e. line-strengths indices) blueward and redward of 5700\,\AA \ come from WHT and GTC data, respectively.

Data reduction was performed using the REDUCEME package \citep{reduceme}, which allows for the careful treatment of error propagation. The reduction process included the standard bias subtraction, flat-fielding, cosmic ray cleaning, distortion (C and S) correction, wavelength calibration, sky subtraction and flux calibration. 

Note that, for our purposes, sky removal is a critical step of the reduction, as both airglow emission and telluric absorption have a prominent role in the raw spectra above $\lambda \sim 7000$\,\AA. To ensure an accurate sky subtraction, an \emph{on-and-off} observational strategy was followed, where the galaxy was offset in the CCD plane during the observations. This allowed for a clean sky emission removal by subtracting consecutive exposures. The second crucial aspect was that of correcting the spectra for telluric absorption features. In order to investigate the impact of such correction on the derived stellar population parameters, we adopted two independent approaches. The spectrophotometric standard star (Hiltner 600, B1) was targeted at the end of the observations. Using the  IRAF task {\it telluric}, the derived atmospheric transmission spectrum was scaled to correct the NGC~1277 data. In addition, we made use of the ESO  tool {\sc Molecfit}~\citep{molecfit}. The latter does not require any extra calibration data (i.e. a spectrophotometric standard star) as it constructs a synthetic atmospheric absorption model for a given spectrum, by fitting  spectral regions dominated by telluric absorption in the spectrum itself\footnote{In particular, we run {\sc Molecfit} by fitting the spectral regions $\lambda \lambda$~8130,8300~\AA; $\lambda \lambda$~7585,7698~\AA; $\lambda \lambda$~7235,7326~\AA \ and $\lambda \lambda$~6861,6924~\AA, where strong H$_2$O and O$_2$ telluric features are found.}. To this effect, molecfit relies on the radiative transfer code {\sc LBLRTM} \citep{Clough}, allowing for a simultaneous fit of both the telluric model and the line spread function to the data. The differences in the inferred stellar population properties among the two approaches are described in Appendix A.

\begin{figure}
\begin{center}
\includegraphics[width=8.cm]{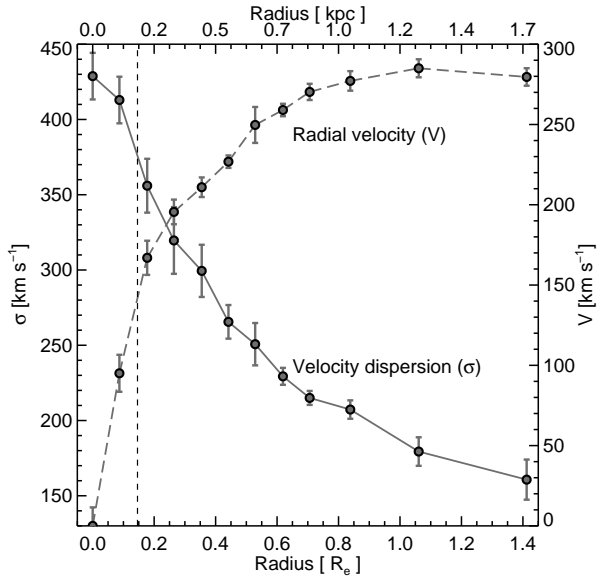}
\end{center}
\caption{Velocity dispersion (solid line, left vertical axis) and radial velocity (dashed line, right vertical axis) profiles as a function of the major-axis radial distance, normalized to the (circularized) effective radius~\re. Dashed vertical line indicates the seeing radius, measured by its average half width at half maximum (HWHM) value (0.4 arcsec).}
\label{fig:kine}
\end{figure}

Since the analysis of stellar populations requires a higher signal-to-noise (SN) than kinematics, we performed a first radial binning by imposing a SN threshold of 20. The radial velocity and velocity dispersion gradients, derived using the public available  software {\sc pPXF}~\citep{ppxf}, are shown in Fig.~\ref{fig:kine}. Our measurements are in good agreement with those published by \citet{remco}, showing that NGC~1277 is pressure-supported in the central regions ($\sigma_c \sim 400$ \kms), whereas ordered motions dominate beyond $\sim0.5$\re. We used the inferred kinematics (radial velocity and velocity dispersion) to correct every row (spectrum) in the image to the rest-frame via linear interpolation of our measurements. Once all the spectra were at rest-frame, we calculated a representative spectrum at each radius of the galaxy with a minimum SN per \AA \ (at~$\lambda \lambda$~6000,6200~\AA) of 80 in the outskirts, achieving over 700 in the central bin. 

\begin{figure*}
\begin{center}
\includegraphics[width=12.cm]{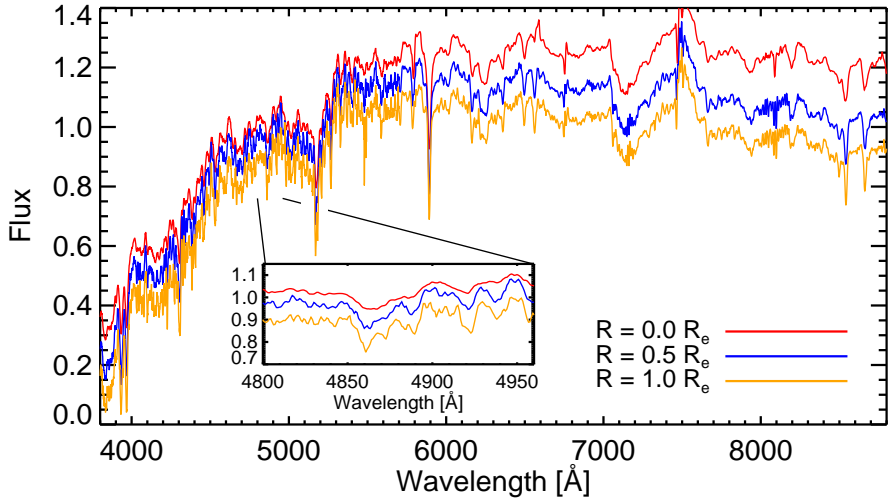}
\end{center}
\caption{Averaged spectra at 0.0 (red), 0.5 (blue) and 1.0 (yellow) \re, arbitrarily scaled in the vertical direction. Note the different spectral resolution between the GTC-OSIRIS ($\lambda > 5700$ \AA) and the WHT ($\lambda < 5700$ \AA) data. Note also the large variation in the velocity dispersion, reflected in a gradual smoothing when moving from the outer 1.0\re \ spectrum to the central radial bin. The inset panel shows a zoom in of the H$\beta$ line.}
\label{fig:spec}
\end{figure*}

Throughout this work, we assume a circularized effective radius of ${\rm R}_{\rm e} = 1.2$~kpc for NGC\,1277, with a 0.344 kpc arcsec$^{-1}$ scale \citep{truji14}. Fig.~\ref{fig:spec} shows three representative spectra at 0, 0.5, and 1 \re , respectively. The change in spectral resolution from the WHT to the OSIRIS data appears at $\lambda = 5700$ \AA. Note also how the intrinsic (associated with the velocity dispersion) resolution varies from the central (in red, $\sigma \sim 400$ \kms) to the outermost bin (in orange, $\sigma \sim 200$ \kms). A zoom around the H$\beta$ feature is shown as an inset, suggesting no significant contamination of the line from nebular emission (see \S~\ref{sec:fitting}).

\section{Analysis and results} \label{sec:fitting}

We based our stellar population analysis on the extended version of the MILES models \citep{miles,miuscat}. The MILES models cover a range from $-2.32$ dex to $+0.22$ dex in total metallicity, and from 0.06 Gyr to 17 Gyr in age. We assumed a bimodal IMF, parametrised as a two-segment function \citep{vazdekis96}, where the star number per logarithmic mass bin is given by:

\begin{equation*}
 \phi(\log m) \mydef \frac{\mathrm{d} \mathcal{N}}{\mathrm{d}\log m } \propto \left\{
        \begin{array}{ll}
            {m_\mathrm{p}^{-\Gamma_\mathrm{b}}} & \quad m \leq 0.2 \, \msun \\
            \\
            p(m) & \quad 0.2 \, \msun < m \leq 0.6 \, \msun \\
            \\
            m^{-\Gamma_\mathrm{b}} & \quad m > 0.6 \, \msun \\
        \end{array}
    \right.
\end{equation*} \label{eq:imf}

\noindent  where $m_\mathrm{p}$ determines the turning-point mass, set to $m_\mathrm{p} = 0.4 \, \msun $ in the \citet{miles,miuscat} models. $p(m)$ is a spline function satisfying a set of boundary conditions so that $\phi(m)$ is continuous:

\begin{eqnarray*}
p(0.2 \, \msun) &=& m_\mathrm{p}^{-\Gamma_\mathrm{b}} \\
p'(0.2 \, \msun) &=& 0 \\
p(0.6 \, \msun) &=& 0.6^{-\Gamma_\mathrm{b}} \\
p'(0.6 \, \msun) &=& -\Gamma_\mathrm{b}  0.6 ^{-(\Gamma_\mathrm{b}+1)}
\end{eqnarray*} \label{eq:bound}

We allowed for a variation in the IMF slope, $\Gamma_\mathrm{b}$, from 0.8 (bottom-light) to 3.3 (bottom-heavy). Under this parametrisation, the standard Kroupa-like IMF is well represented by a slope of  $\Gamma_\mathrm{b} = 1.3$. Note that $\Gamma_{\rm b}$ varies the slope of the high-mass end of the (bimodal) IMF, in contrast to other studies where the high-mass end of the IMF is kept fixed, while the low-mass end is varied \citep[e.g.][]{cvd12}. Since the (bimodal) IMF is normalized to have a total mass of 1~$M_\odot$,  changing $\Gamma_{\rm b}$ does actually vary  the mass fraction of low- to high-mass stars. Hence, for the present study, $\Gamma_{\rm b}$ should not be regarded as the high-mass end slope of the IMF, but rather as a proxy for the fraction of low-mass stars.

\begin{figure}
\begin{center}
\includegraphics[width=8.cm]{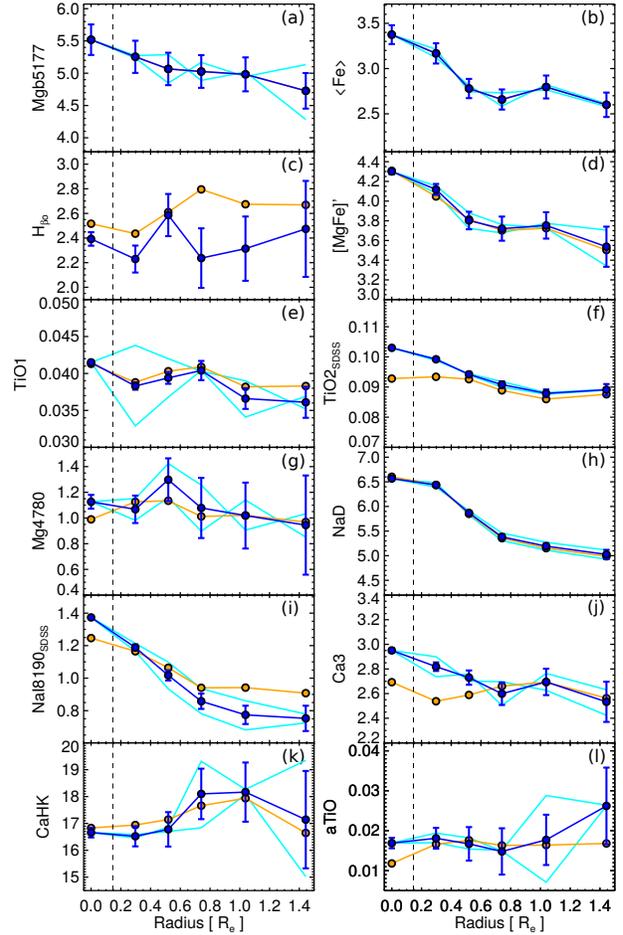}
\end{center}
\caption{Observed gradients of the age-, metallicity- and IMF-sensitive indices considered in the fitting process (blue symbols). Dark blue represents the mean radial profile, whereas in light blue we show the radial gradient at each side of the galaxy. The 1$\sigma$ error bars are also shown for the mean profile. The best-fitting solution is overplotted in orange. The definition of all spectral indices are given in \citet{labarbera}, while aTiO has been computed according to the definition of \citet{Spiniello2013}. Although not individually included in the fitting process, panels (a) and (b) show the Mgb and $\langle$Fe$\rangle$ line-strength profiles. Note that, according to the MILES models, both indices have quite similar sensitivities to a variation in total metallicity. Since they also exhibit the same radial variation ($\sim 0.8$\AA), a flat [Mg/Fe] gradient is expected, consistent with Fig.~\ref{fig:pob}. In all panels, dashed vertical lines mark the seeing radius.}
\label{fig:ind}
\end{figure}

To determine the stellar population parameters we followed the same approach as in \citet{labarbera}. Age, metallicity, and IMF slope were obtained by minimising the following equation:

\begin{equation}
\chi^2(\Gamma_\mathrm{b}, \mathrm{age}, {\rm [Z/H]})= 
  \sum_i \left[ \frac{(EW_i-\Delta_{\alpha,i}) - EW_{M,i} }{\sigma_{EW_i}} \right]^2,
\end{equation} \label{eq:method}

\noindent where the sum expands over the $i$th observed and predicted line-strength indices included in the analysis. $EW_i$ are the measured line-strengths, $EW_{M,i}$ the predictions from MILES models, and $\Delta_{\alpha,i}$ is a  correction applied to each index to account for deviations from a non-solar abundance scale, and the effect of ``residual'' abundance ratios on line strengths (see \citealt{labarbera} and Appendix~A, for details). To derive  [Mg/Fe], for each radial bin we estimated separately the Magnesium and the Iron metallicities (Z$_\mathrm{Mg}$ and Z$_\mathrm{Fe}$, respectively) through index-index fitting of the H$_{\beta_\mathrm{o}}$ vs. Mgb and $\langle$Fe$\rangle$ line-strengths, respectively. We then subtracted one metallicity from the other to calculate the [Z$_\mathrm{Mg}$/Z$_\mathrm{Fe}$] proxy, which tightly correlates with the ``true'' [Mg/Fe] \citep[see Fig.6 in][]{labarbera}. In Fig.~\ref{fig:ind} we show the measured radial profiles of all  indices considered in the fitting process, along with the best-fitting SSP solution at each radius. Notice that the H$_{\beta_\mathrm{o}}$ line has been corrected for nebular emission contamination, as detailed in Appendix~A.

\begin{figure}
\begin{center}
\includegraphics[width=7.5cm]{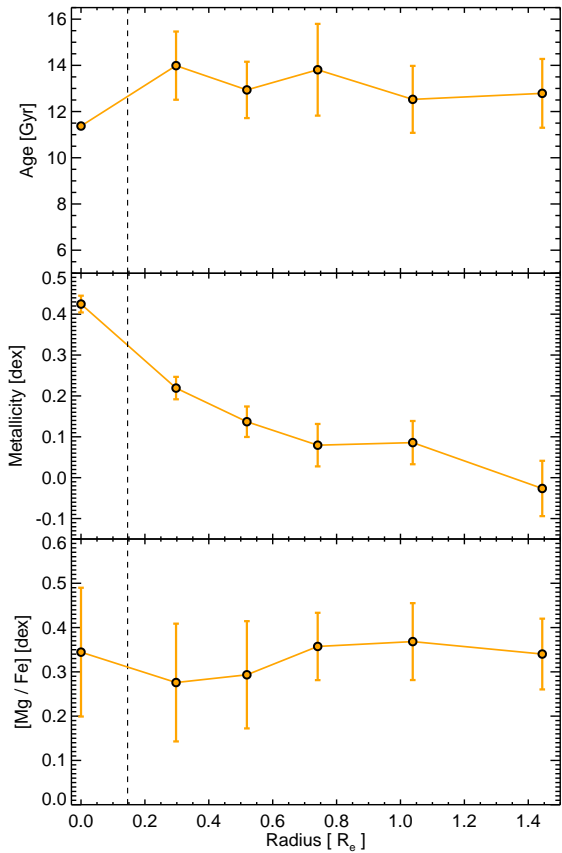}
\end{center} 
\caption{Best-fitting age (top), metallicity (middle) and [Mg/Fe] (bottom) gradients. Both age and [Mg/Fe] show a fairly flat radial behaviour, whereas  metallicity decreases by 0.4 dex from the centre outwards to 1.5\re.  An extrapolation of stellar population models is required to account for line-strength values in the central region. In all panels, dashed vertical lines mark the seeing radius.}
\label{fig:pob}
\end{figure}

The derived age, metallicity and [Mg/Fe] radial profiles are shown in Fig.~\ref{fig:pob}. NGC~1277 exhibits a flat age profile, always above 10 Gyrs. The metallicity, on the contrary, decreases with increasing radius. Notice that the metallicity in the central bin (+0.41 dex) exceeds the maximum value of MILES models (+0.22). Hence, for this specific bin, we extrapolated the models  \citep[in the same way as in][]{labarbera} to match the observed indices (in particular, the total metallicity indicator $\rm [MgFe]'$). Although in general there is a good agreement between the observed and best-fitting line strengths in Fig.~\ref{fig:ind}, some discrepancies exist. In particular, the mismatch between observed and model indices in the most central bins (see, e.g., panels (f) and (i)), can be at least partly understood as a limitation of the stellar population models to describe very metal-rich ($\gtrsim 0.2$ dex) populations and/or uncertainties in the age estimate. However, none of these issues significantly affect  the results of our work, as discussed in Appendix~A. Fig.~\ref{fig:pob} also shows that the [Mg/Fe] gradient of NGC\,1277 is flat, with [Mg/Fe]$\sim$ 0.3 dex at all radii. This flat [Mg/Fe] radial profile can be also directly inferred from the fact that 
both the Mgb and $\langle$Fe$\rangle$ spectral indices (see panels (a) and (b) of Fig.~\ref{fig:ind}) show a similar amount of radial variation~\footnote{Notice, in fact, that in MILES models, both Mgb and $\langle$Fe$\rangle$ have a similar sensitivity to total metallicity, hence a similar radial gradient for the two indices implies a flat trend of the [Z$_\mathrm{Mg}$/Z$_\mathrm{Fe}$] proxy.}. Note that the large error bars on [Mg/Fe] in the central regions reflect the uncertainties on the [Fe/H] and [Mg/H] determination/extrapolation used to derive 
the [Mg/Fe] proxy. The gradients shown in Fig.~\ref{fig:pob} are in agreement with those published by \citet[][Fig.~4 in their paper]{truji14}. This is not a trivial result, since here we have allowed for possible  radial variations of the IMF.

\begin{figure}
\begin{center}
\includegraphics[width=8.cm]{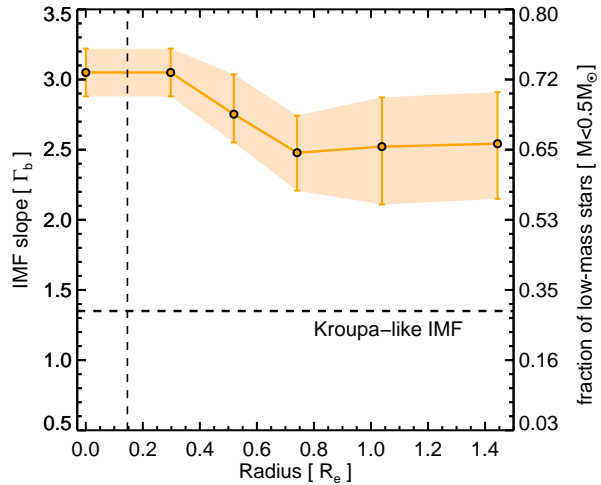}
\end{center}
\caption{Radial IMF-slope gradient for the relic massive galaxy NGC\,1277. The left vertical axis indicates the slope of the IMF in the bimodal case ($\Gamma_\mathrm{b}$), i.e., the slope of the IMF for stars with masses larger than 0.5\msun. This $\Gamma_\mathrm{b}$ parameter can be transformed into a dwarf-to-giant ratio (right vertical axis), which is what the line-strength indices are actually tracing \citep{labarbera}. For reference, the Milky-Way IMF value is shown as a horizontal dashed line. The inferred IMF gradient is mildly decreasing and it is very bottom-heavy up to 1.5\re. Uncertainties (the orange shaded region) are obtained by removing  line-strength indices from the fitting process which might be potentially affected by flux calibration, telluric correction or modelling uncertainties. Dashed vertical line indicates the extent of the seeing disc.}
\label{fig:imf}
\end{figure}

The IMF radial profile of NGC~1277 is shown in Fig.~\ref{fig:imf}, where the shaded region accounts for the scatter among $\Delta\Gamma_\mathrm{b} $ values from different tests, where the fitting process has been repeated with different sets of spectral indices and different methods to remove telluric lines as detailed in Appendix~A. Regarding the discrepancies between the observed and best-fitting line strengths (see Fig.~\ref{fig:ind}), note  that the TiO$_2$ and NaI\,8190 lines can be potentially affected by errors in the flux calibration (the former) and telluric residuals  (both).  Although the central TiO$_2$ value might suggest a slightly steeper IMF profile, our best-fitting solution points to a mild radial variation of IMF slope ($\Gamma_\mathrm{b} \sim -0.5$), i.e. a bottom-heavy IMF at all radii. Notice that in the present analysis we have not included the CaH1\footnote{Although not used in our analysis, the radial variation of the CaH1 index is mild (0.002 mag) further supports a rather flat IMF profile.} and CaH2 IMF-sensitive features of \citet{Spiniello2014}, as (1) CaH1 has been shown to be not matched by extended MILES models \citep[regardless of the IMF, see][]{labarbera}, while (2) CaH2 is strongly affected by the O$_2$ ($B$-band) atmospheric absorption at $\lambda$~$\sim$~6900~\AA \, in our data. Concerning Ca3, the (observed) index tends to decrease with radius, while our best-fitting solutions favour a flat radial behaviour for Ca3.  This discrepancy might be related to some radial variation of the [Ca/Fe] abundance ratio which is not well captured by our fitting approach. In fact, although we have included  [Ca/Fe] as a free fitting parameter (see Appendix~A), the [Ca/Fe] is mainly constrained, in this work, from the CaHK line, which is strongly affected by the abundance pattern of other alpha elements \citep[e.g.][]{alpha}. Nevertheless, a decreasing trend of Ca3 with radius is the opposite behaviour to what one would expect for a less bottom-heavy IMF at larger radii (as Ca3 decreases with $\Gamma_\mathrm{b}$). This is consistent with our conclusion that the IMF of NGC1277 shows only a mild radial variation. In fact, as shown in Appendix~A, removing this index altogether from the fits does not significantly change the radial trend of  $\Gamma_\mathrm{b}$. Regarding aTiO, the only point that deviates significantly from the best-fits is the innermost one. This could be related to some issue in the extrapolation of the models at high metallicity, and/or some peculiar abundance ratio for R=0\,\re. However, as the deviation affects only one single point, it does not change our general conclusions about the radial behaviour of $\Gamma_\mathrm{b}$ for NGC1277.

The robustness of the IMF profile, against all issues mentioned so far, is demonstrated in Appendix~A, where  we show how the mildly decreasing behaviour of $\Gamma_\mathrm{b}$ persists under different telluric correction procedures (see \S~\ref{sec:data}), as well as different combinations of indices included in the fits, and changes in the modelling assumptions.

Ultimately, gravity-sensitive indices trace the dwarf-to-giant star ratio in the IMF, regardless of the adopted IMF parametrization \citep{labarbera}. This is explicitly shown in the right vertical axis of Fig.~\ref{fig:imf}, where we represent the fraction (in mass) of stars below 0.5M$_\odot$ for each IMF slope. The inferred dwarf-to-giant ratio in NGC~1277 is, at all radii, at least a factor of 2 larger than in the Milky Way (horizontal dashed line). 

\begin{figure}
\begin{center}
\includegraphics[width=8.cm]{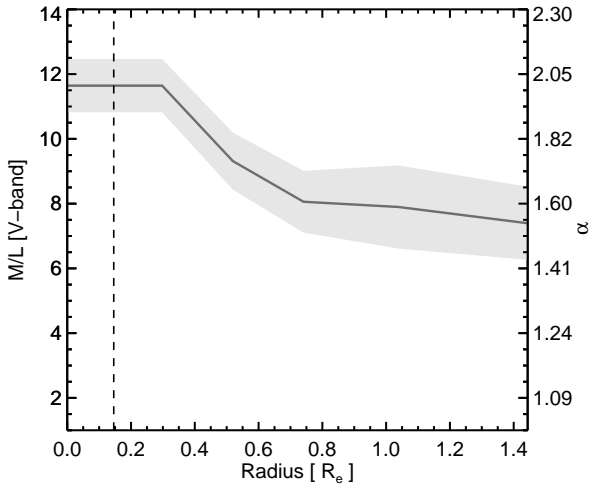}
\end{center}
\caption{Radial M/L ratio ($V$-band) gradient for the massive compact galaxy NGC\,1277, under the assumption of a bimodal IMF. At each radius, the best-fitting age, metallicity and IMF slope shown in Figs.~\ref{fig:pob} and \ref{fig:imf} translate into a $V$-band M/L ratio.  The right-hand axis represents the ``mismatch parameter'', $\alpha$, defined as the ratio between the inferred M/L and that expected from a stellar population with the same age and metallicity as observed, but with a Milky Way-like IMF slope. Notice that the absolute values of the M/L gradients heavily depend on the parametrisation adopted for the IMF \citep{ferreras14}. The light shaded region accounts for the uncertainties in all the stellar population parameters (e.g., error bars in Fig.~\ref{fig:pob} and shaded region in Fig.~\ref{fig:imf}). Dashed vertical lines mark the seeing radius.}
\label{fig:ml}
\end{figure}

The inferred stellar population properties of NGC\,1277 (ages, metallicities, and IMF slopes; as shown in Figs.~\ref{fig:pob} and \ref{fig:imf}) can be used to estimate the stellar M/L radial profile. Notice, however, that for old stellar systems ($\gtrsim$ 10 Gyr), stellar population analysis cannot provide fully reliable mass-to-light ratios, as it relies on the assumption of a given  parametrisation for the IMF\citep{ferreras14}. In this sense, our favoured bimodal distribution seems to better match the dynamical M/L estimates \citep{labarbera,Spiniello2013}, while a single power-law distribution provides  unrealistically high M/L's. For a bimodal IMF parametrisation, the expected M/L radial profile of NGC\,1277 is shown in Fig.~\ref{fig:ml}.

\section{Discussion} \label{sec:discussion}

\subsection{The excess of low-mass stars in ETGs}
What drives the excess of low-mass stars in ETGs remains unknown, however there are presently two distinct candidates, namely [Mg/Fe] and $\sigma$. The latter may be viewed as a dynamical controlling parameter, whereas the former is a stellar population parameter. 

\begin{figure}
\begin{center}
\includegraphics[width=8.cm]{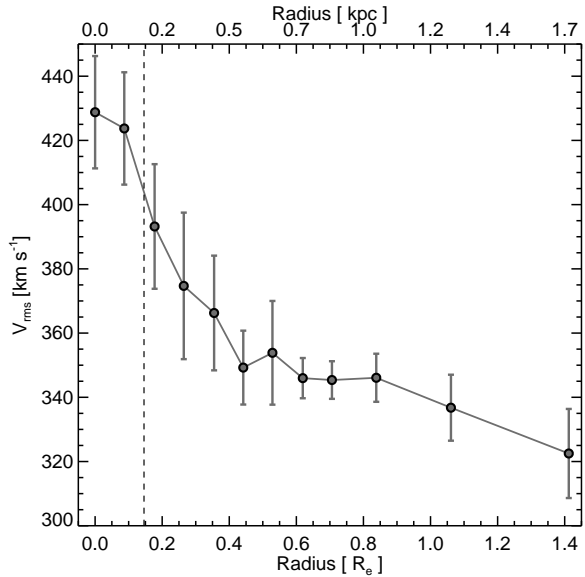}
\end{center}
\caption{Radial V$_\mathrm{rms}$ profile of NGC\,1277. This quantity, accounting for the total kinetical energy, is more closely related to the effective velocity dispersion, which has been found to tightly correlate with the observed IMF variations. The radial V$_\mathrm{rms}$ variation is significant, albeit milder, than the $\sigma$ variation. Dashed vertical lines mark the seeing radius.}
\label{fig:v05}
\end{figure}

Dynamical studies \citep{treu,cappellari} suggest that the galaxy central velocity dispersion 
correlates strongly with the IMF slope - with more massive galaxies having steeper IMF slopes than less massive galaxies. This result has found support from several groups using different methodologies \citep{ferreras,labarbera,Spiniello2013}. 
However, based on our analysis, we {\it reject the hypothesis that local velocity dispersion is the principal agent driving IMF variations}. Given the velocity dispersion gradient observed in NGC~1277 (from~$\sim$~420~\kms to 150 \kms), if the correlation between IMF slope and central velocity dispersion holds also for local velocity dispersion, we would expect a typical radial change in the IMF slope of $\Delta\Gamma_\mathrm{b}= -2.4$. This value is incompatible with our inferred IMF gradient ($\Delta\Gamma_\mathrm{b} \sim -0.5$). Note, however, that our conclusions apply to the local velocity dispersion, not to the effective central velocity dispersion. To investigate the relation between IMF slope and local kinematical properties of NGC\,1277, we show in Fig.~\ref{fig:v05} its V$_\mathrm{rms}$ radial profile, defined as:

\begin{equation*}
V_\mathrm{rms} = \sqrt{V_\mathrm{rot}^2 + \sigma^2}
\end{equation*} \label{eq:so5}

\noindent Note that the effective velocity dispersion analysed in \citet{labarbera} and in \citet{Spiniello2013}, which was estimated using a 3 arcsec wide fibre, is better represented by V$_\mathrm{rms}$ rather than by the local sigma, since the latter does not include the line broadening associated with the rotational velocity profile. Both $\Gamma_\mathrm{b}$ and V$_\mathrm{rms}$ decrease with radius; V$_\mathrm{rms}$ is greater than 300\kms, and $\Gamma_\mathrm{b}$ is steeper than the Milky Way value at all radii in NGC~1277. This is consistent with previous works reporting a bottom-heavy IMF at high {\it effective}  velocity dispersion.

As an alternative explanation for the IMF variations, based on stellar population analysis, \citet{cvd12} found that the IMF slope also correlates with the overabundance of $\alpha$-elements, in particular with the [Mg/Fe] ratio, in the sense that galaxies more enhanced in [Mg/Fe] have steeper IMF slopes.
As we mentioned in \S\ref{sec:fitting}, our [Mg/Fe] measurements are heavily dependent on the interpolation scheme we use, and we find that our data can entertain either slightly decreasing or increasing [Mg/Fe] radial gradients. Therefore, from the present study we cannot assess the relation between IMF slope and [Mg/Fe] \citep[but see][]{LB15}.

To understand the differences among the aforementioned studies, \citet{smith} has independently reviewed the dynamical results from \citet{cappellari} and those based on stellar population analysis of \citet{cvd12}. Although he found an overall agreement between both studies, the galaxy-by-galaxy comparison shows that the correlation between IMF slope and velocity dispersion does not agree with that found when comparing the IMF and [Mg/Fe].

The rapidly growing observational evidence of a variable IMF slope may be in tension with other well-established properties of ETGs in the local Universe. In particular, chemical evolution models assuming a bottom-heavy IMF are unable to reproduce the observed metallicities and abundance ratios of ETGs \citep{Arrigoni}. This idea has been recently revisited in a series of articles by \citet{weidner:13} and \citet{Ferreras15} (see also \citealt{vazdekis:97,Larson}) who found that a time-dependent IMF is necessary to explain the stellar population properties of massive ETGs at $z\sim0$. Note that, according to \citet{Ferreras15}, a constant-in-time IMF can be ruled out, not only for a bimodal IMF shape, but also for a two-segment IMF parametrization \citep[e.g. as defined in][]{cvd12}, and even for a V-shaped IMF (characterized by an enhanced fraction of both low- and high-mass stars).

\subsection{The emerging picture}
Our analysis of the stellar population properties of NGC~1277 is summarized in Figs.~\ref{fig:pob} and  \ref{fig:imf}. The radial age profile of this galaxy, even allowing for possible IMF variations, indicates that the underlying stellar population is very old ($\gtrsim 10$ Gyr) at all radii, with no evidence of recent star formation \citep{truji14}. This result suggests that the stellar component of NGC~1277 was formed at high redshift ($z \gtrsim 2$), and that solving for the IMF at each radius does not convert NGC~1277 into a young, massive and compact system, as those identified in the local Universe~\citep{truji09,anna12,anna13}.

In terms of kinematics, NGC~1277 shows an extreme radial velocity dispersion gradient, ranging from more than 400 \kms in the centre to $\sigma~\sim$~150 \kms  beyond 1~\re. The rotation in this galaxy shows the opposite behaviour, rising to $V_\mathrm{rot} \sim 280$ \kms at 1~\re. Interestingly, these differences in the kinematical properties between the centre and the outskirts of NGC~1277 are not reflected in the properties of its stellar populations. The radial variation of $\alpha$-element overabundance, age and IMF slope are mild, whereas the total metallicity decreases by $\sim$ 0.4 dex from the centre to 1~\re. These gradients are consistent with a \textit{fast, monolithic-like formation process}~\citep{pat}. 
In this sense, in terms of its stellar population properties, NGC~1277 does not differ from typical massive ETGs in the Local Universe. However, its extreme dynamics and compact morphology set it apart from the vast majority of local ETGs.

The IMF radial profile of NGC~1277 further constrains the formation process of massive ETGs. NGC~1277 shows much weaker radial variations of the IMF slope ($\Delta\Gamma_\mathrm{b}= -0.5$) when compared to the IMF gradient of the two massive galaxies analysed by \citet{mn14} ($\Delta\Gamma_\mathrm{b} = -1.5$ ). Our suggestion is that NGC~1277 represents the monolithic-like formation phase of massive ETGs. In this picture the later accretion of less massive satellites (with ``standard'' Kroupa-like IMF slopes), which would settle preferentially in the galaxy periphery due to their lower densities, would steepen the pristine IMF gradient making it more similar to that found in nearby massive galaxies~\citet{mn14} . 

Indeed, we hypothesize that relic galaxies, such as NGC~1277, will evolve to become the cores of massive ellipticals in the nearby Universe~\citep{truji14}. The bulk of the stellar mass of this galaxy was formed in a short ($\tau \sim 100$ Myr) and intense (SFR $\gtrsim 10^{3}$ \msun/yr) burst at redshift $z \gtrsim 2$, but remained concentrated within a few kpcs in the centre of the dark matter halo. Afterwards, through dry merging \citep{vdk,truji11,ferreras14}, the system would typically evolve to the characteristic sizes and stellar masses that are observed at $z \sim 0$. This interpretation is supported by numerical simulations \citep[e.e.][]{oser,navarro}, and also by stellar population studies of the outskirts of nearby ETGs \citep{coccato,flb12,pasto,montes}. Since dry mergers cannot significantly alter the properties of the stellar populations, a relic massive galaxy should present the same stellar population properties as the core of a typical $z \sim 0$, massive ETG, but with the latter's morphology and kinematics shaped by successive mergers.

It may seem surprising, at face value, that such a relic system should be identified in such a dense and dynamic environment such as that represented by the Perseus cluster \citep[$M_{200} = 7.74 \times 10^{14}$\msun,][]{perseus}. However, this is in fact a natural expectation in the hierarchical clustering paradigm since overdensities in what are now galaxy clusters were the first structures to collapse and to form massive galaxies \citep{Mo}. In addition, the active intracluster environment may have inhibited the subsequent  formation of younger stellar populations within NGC~1277 either via the stripping or heating of cold gas \citep{edge,odea}. In this sense, \citet{Martin} have recently shown that in the BOLSHOI numerical simulation \citep{bolshoi}, dark matter haloes similar to those hosting NGC~1277-like objects have experienced below-average mass accretion, or even mass loss, since $z=2$.

Finally, we find that the radial IMF behaviour seen in NGC~1277 is not driven by the local velocity dispersion, and therefore, it cannot explain the IMF variations seen in ETGs. The fact that the IMF slope can vary radially within galaxies \citep{mn14} suggests a complex galaxy formation scenario, with significantly different enrichment and supernovae-feedback rates when comparing the centre and the outskirts of ETGs. Thus, more observational efforts are needed to better understand the IMF behaviour in nearby galaxies, investigating what stellar population properties can explain the observed IMF variations. In this sense, large resolved spectroscopic surveys such as CALIFA \citep{califa} and MaNGA will provide statistically significant samples. In addition, the joint analysis of the IMF inferred from dynamical modelling and from stellar population analysis can reveal important information about the interplay between galaxy stellar population properties and dynamics. The intrinsic degeneracies of these methods ($M/L$ ratios, dark matter content, stellar remnants and central black-hole mass) may be broken if both approaches are combined.

\vspace{0.2in}

\small{ \footnotesize{\textit{\textbf{Acknowledgements}}} We thank Ignacio Ferreras for his insightful comments and suggestions on the paper. Also, we would like to thank the two anonymous referees for their precise and constructive suggestions. We acknowledge support from the Spanish Government grant AYA2013-48226-C3-1-P. Based on observations made with the WHT and the GTC telescopes, both installed in the Spanish Observatorio del Roque de los Muchachos of the Instituto de Astrof\'\i sica de Canarias.}


\begin{thebibliography}{56}
\expandafter\ifx\csname natexlab\endcsname\relax\def\natexlab#1{#1}\fi

\bibitem[{{Aleksi{\'c}} {et~al}\mbox{.}(2010){Aleksi{\'c}}, {Antonelli},
  {Antoranz}, {Backes}, {Baixeras}, {Balestra}, {Barrio}, {Bastieri}, {Becerra
  Gonz{\'a}lez}, {Bednarek}, {Berdyugin}, {Berger}, {Bernardini}, {Biland},
  {Bock}, {Bonnoli}, {Bordas}, {Borla Tridon}, {Bosch-Ramon}, {Bose}, {Braun},
  {Bretz}, {Britzger}, {Camara}, {Carmona}, {Carosi}, {Colin}, {Commichau},
  {Contreras}, {Cortina}, {Costado}, {Covino}, {Dazzi}, {De Angelis}, {De Cea
  del Pozo}, {De los Reyes}, {De Lotto}, {De Maria}, {De Sabata}, {Delgado
  Mendez}, {Doert}, {Dom{\'{\i}}nguez}, {Dominis Prester}, {Dorner}, {Doro},
  {Elsaesser}, {Errando}, {Ferenc}, {Fonseca}, {Font}, {Galante},
  {Garc{\'{\i}}a L{\'o}pez}, {Garczarczyk}, {Gaug}, {Godinovic}, {Hadasch},
  {Herrero}, {Hildebrand}, {H{\"o}hne-M{\"o}nch}, {Hose}, {Hrupec}, {Hsu},
  {Jogler}, {Klepser}, {Kr{\"a}henb{\"u}hl}, {Kranich}, {La Barbera}, {Laille},
  {Leonardo}, {Lindfors}, {Lombardi}, {Longo}, {L{\'o}pez}, {Lorenz},
  {Majumdar}, {Maneva}, {Mankuzhiyil}, {Mannheim}, {Maraschi}, {Mariotti},
  {Mart{\'{\i}}nez}, {Mazin}, {Meucci}, {Miranda}, {Mirzoyan}, {Miyamoto},
  {Mold{\'o}n}, {Moles}, {Moralejo}, {Nieto}, {Nilsson}, {Ninkovic}, {Orito},
  {Oya}, {Paiano}, {Paoletti}, {Paredes}, {Partini}, {Pasanen}, {Pascoli},
  {Pauss}, {Pegna}, {Perez-Torres}, {Persic}, {Peruzzo}, {Prada}, {Prandini},
  {Puchades}, {Puljak}, {Reichardt}, {Rhode}, {Rib{\'o}}, {Rico}, {Rissi},
  {R{\"u}gamer}, {Saggion}, {Saito}, {Salvati}, {S{\'a}nchez-Conde},
  {Satalecka}, {Scalzotto}, {Scapin}, {Schultz}, {Schweizer}, {Shayduk},
  {Shore}, {Sierpowska-Bartosik}, {Sillanp{\"a}{\"a}}, {Sitarek}, {Sobczynska},
  {Spanier}, {Spiro}, {Stamerra}, {Steinke}, {Struebig}, {Suric}, {Takalo},
  {Tavecchio}, {Temnikov}, {Terzic}, {Tescaro}, {Teshima}, {Torres}, {Vankov},
  {Wagner}, {Zabalza}, {Zandanel}, {Zanin}, {Zapatero}, {Pfrommer}, {Pinzke},
  {En{\ss}lin}, {Inoue}, {Ghisellini}, \& {MAGIC Collaboration}}]{perseus}
{Aleksi{\'c}} J. {et~al.}, 2010, \apj, 710, 634

\bibitem[{{Arrigoni} {et~al}\mbox{.}(2010){Arrigoni}, {Trager}, {Somerville},
  \& {Gibson}}]{Arrigoni}
{Arrigoni} M., {Trager} S.~C., {Somerville} R.~S., {Gibson} B.~K., 2010,
  \mnras, 402, 173

\bibitem[{{Cappellari} \& {Emsellem}(2004)}]{ppxf}
{Cappellari} M., {Emsellem} E., 2004, \pasp, 116, 138

\bibitem[{{Cappellari} {et~al}\mbox{.}(2012){Cappellari}, {McDermid},
  {Alatalo}, {Blitz}, {Bois}, {Bournaud}, {Bureau}, {Crocker}, {Davies},
  {Davis}, {de Zeeuw}, {Duc}, {Emsellem}, {Khochfar}, {Krajnovi{\'c}},
  {Kuntschner}, {Lablanche}, {Morganti}, {Naab}, {Oosterloo}, {Sarzi}, {Scott},
  {Serra}, {Weijmans}, \& {Young}}]{cappellari}
{Cappellari} M. {et~al.}, 2012, \nat, 484, 485

\bibitem[{{Cardiel}(1999)}]{reduceme}
{Cardiel} N., 1999

\bibitem[{{Cenarro} {et~al}\mbox{.}(2001){Cenarro}, {Cardiel}, {Gorgas},
  {Peletier}, {Vazdekis}, \& {Prada}}]{cat}
{Cenarro} A.~J., {Cardiel} N., {Gorgas} J., {Peletier} R.~F., {Vazdekis} A.,
  {Prada} F., 2001, \mnras, 326, 959

\bibitem[{{Cenarro} {et~al}\mbox{.}(2003){Cenarro}, {Gorgas}, {Vazdekis},
  {Cardiel}, \& {Peletier}}]{cenarro}
{Cenarro} A.~J., {Gorgas} J., {Vazdekis} A., {Cardiel} N., {Peletier} R.~F.,
  2003, \mnras, 339, L12

\bibitem[{{Cervantes} \& {Vazdekis}(2009)}]{Cervantes}
{Cervantes} J.~L., {Vazdekis} A., 2009, \mnras, 392, 691

\bibitem[{{Chabrier}(2003)}]{chabrier}
{Chabrier} G., 2003, \pasp, 115, 763

\bibitem[{{Clough} {et~al}\mbox{.}(2005){Clough}, {Shephard}, {Mlawer},
  {Delamere}, {Iacono}, {Cady-Pereira}, {Boukabara}, \& {Brown}}]{Clough}
{Clough} S.~A., {Shephard} M.~W., {Mlawer} E.~J., {Delamere} J.~S., {Iacono}
  M.~J., {Cady-Pereira} K., {Boukabara} S., {Brown} P.~D., 2005, \jqsrt, 91,
  233

\bibitem[{{Coccato} {et~al}\mbox{.}(2010){Coccato}, {Gerhard}, \&
  {Arnaboldi}}]{coccato}
{Coccato} L., {Gerhard} O., {Arnaboldi} M., 2010, \mnras, 407, L26

\bibitem[{{Conroy} \& {van Dokkum}(2012)}]{cvd12}
{Conroy} C., {van Dokkum} P.~G., 2012, \apj, 760, 71

\bibitem[{{Edge} {et~al}\mbox{.}(1992){Edge}, {Stewart}, \& {Fabian}}]{edge}
{Edge} A.~C., {Stewart} G.~C., {Fabian} A.~C., 1992, \mnras, 258, 177

\bibitem[{{Emsellem}(2013)}]{Emsellem}
{Emsellem} E., 2013, \mnras, 433, 1862

\bibitem[{{Ferr{\'e}-Mateu} {et~al}\mbox{.}(2013){Ferr{\'e}-Mateu}, {Vazdekis},
  \& {de la Rosa}}]{anna13}
{Ferr{\'e}-Mateu} A., {Vazdekis} A., {de la Rosa} I.~G., 2013, \mnras, 431, 440

\bibitem[{{Ferr{\'e}-Mateu} {et~al}\mbox{.}(2012){Ferr{\'e}-Mateu}, {Vazdekis},
  {Trujillo}, {S{\'a}nchez-Bl{\'a}zquez}, {Ricciardelli}, \& {de la
  Rosa}}]{anna12}
{Ferr{\'e}-Mateu} A., {Vazdekis} A., {Trujillo} I., {S{\'a}nchez-Bl{\'a}zquez}
  P., {Ricciardelli} E., {de la Rosa} I.~G., 2012, \mnras, 423, 632

\bibitem[{{Ferreras} {et~al}\mbox{.}(2013){Ferreras}, {La Barbera}, {de la
  Rosa}, {Vazdekis}, {de Carvalho}, {Falc{\'o}n-Barroso}, \&
  {Ricciardelli}}]{ferreras}
{Ferreras} I., {La Barbera} F., {de la Rosa} I.~G., {Vazdekis} A., {de
  Carvalho} R.~R., {Falc{\'o}n-Barroso} J., {Ricciardelli} E., 2013, \mnras,
  429, L15

\bibitem[{{Ferreras} {et~al}\mbox{.}(2014){Ferreras}, {Trujillo},
  {M{\'a}rmol-Queralt{\'o}}, {P{\'e}rez-Gonz{\'a}lez}, {Cava}, {Barro},
  {Cenarro}, {Hern{\'a}n-Caballero}, {Cardiel},
  {Rodr{\'{\i}}guez-Zaur{\'{\i}}n}, \& {Cebri{\'a}n}}]{ferreras14}
{Ferreras} I. {et~al.}, 2014, \mnras, 444, 906

\bibitem[{{Ferreras} {et~al}\mbox{.}(2015){Ferreras}, {Weidner}, {Vazdekis}, \&
  {La Barbera}}]{Ferreras15}
{Ferreras} I., {Weidner} C., {Vazdekis} A., {La Barbera} F., 2015, \mnras, 448,
  L82

\bibitem[{{Kausch} {et~al}\mbox{.}(2014){Kausch}, {Noll}, {Smette},
  {Kimeswenger}, {Horst}, {Sana}, {Jones}, {Barden}, {Szyszka}, \&
  {Vinther}}]{molecfit}
{Kausch} W. {et~al.}, 2014, in Astronomical Society of the Pacific Conference
  Series, Vol. 485, Astronomical Data Analysis Software and Systems XXIII,
  {Manset} N., {Forshay} P., eds., p. 403

\bibitem[{{Klypin} {et~al}\mbox{.}(2011){Klypin}, {Trujillo-Gomez}, \&
  {Primack}}]{bolshoi}
{Klypin} A.~A., {Trujillo-Gomez} S., {Primack} J., 2011, \apj, 740, 102

\bibitem[{{Kroupa}(2002)}]{kroupa}
{Kroupa} P., 2002, Science, 295, 82

\bibitem[{{La Barbera} {et~al}\mbox{.}(2012){La Barbera}, {Ferreras}, {de
  Carvalho}, {Bruzual}, {Charlot}, {Pasquali}, \& {Merlin}}]{flb12}
{La Barbera} F., {Ferreras} I., {de Carvalho} R.~R., {Bruzual} G., {Charlot}
  S., {Pasquali} A., {Merlin} E., 2012, \mnras, 426, 2300

\bibitem[{{La Barbera} {et~al}\mbox{.}(2015){La Barbera}, {Ferreras}, \&
  {Vazdekis}}]{LB15}
{La Barbera} F., {Ferreras} I., {Vazdekis} A., 2015, \mnras, 449, L137

\bibitem[{{La Barbera} {et~al}\mbox{.}(2013){La Barbera}, {Ferreras},
  {Vazdekis}, {de la Rosa}, {de Carvalho}, {Trevisan}, {Falc{\'o}n-Barroso}, \&
  {Ricciardelli}}]{labarbera}
{La Barbera} F., {Ferreras} I., {Vazdekis} A., {de la Rosa} I.~G., {de
  Carvalho} R.~R., {Trevisan} M., {Falc{\'o}n-Barroso} J., {Ricciardelli} E.,
  2013, \mnras, 433, 3017

\bibitem[{{Larson}(1998)}]{Larson}
{Larson} R.~B., 1998, \mnras, 301, 569

\bibitem[{{Mart{\'{\i}}n-Navarro} {et~al}\mbox{.}(2015){Mart{\'{\i}}n-Navarro},
  {Barbera}, {Vazdekis}, {Falc{\'o}n-Barroso}, \& {Ferreras}}]{mn14}
{Mart{\'{\i}}n-Navarro} I., {Barbera} F.~L., {Vazdekis} A.,
  {Falc{\'o}n-Barroso} J., {Ferreras} I., 2015, \mnras, 447, 1033

\bibitem[{{Mo} \& {White}(1996)}]{Mo}
{Mo} H.~J., {White} S.~D.~M., 1996, \mnras, 282, 347

\bibitem[{{Montes} {et~al}\mbox{.}(2014){Montes}, {Trujillo}, {Prieto}, \&
  {Acosta-Pulido}}]{montes}
{Montes} M., {Trujillo} I., {Prieto} M.~A., {Acosta-Pulido} J.~A., 2014,
  \mnras, 439, 990

\bibitem[{{Naab} {et~al}\mbox{.}(2009){Naab}, {Johansson}, \&
  {Ostriker}}]{naab}
{Naab} T., {Johansson} P.~H., {Ostriker} J.~P., 2009, \apjl, 699, L178

\bibitem[{{Navarro-Gonz{\'a}lez} {et~al}\mbox{.}(2013){Navarro-Gonz{\'a}lez},
  {Ricciardelli}, {Quilis}, \& {Vazdekis}}]{navarro}
{Navarro-Gonz{\'a}lez} J., {Ricciardelli} E., {Quilis} V., {Vazdekis} A., 2013,
  \mnras, 436, 3507

\bibitem[{{O'Dea} {et~al}\mbox{.}(2008){O'Dea}, {Baum}, {Privon}, {Noel-Storr},
  {Quillen}, {Zufelt}, {Park}, {Edge}, {Russell}, {Fabian}, {Donahue},
  {Sarazin}, {McNamara}, {Bregman}, \& {Egami}}]{odea}
{O'Dea} C.~P. {et~al.}, 2008, \apj, 681, 1035

\bibitem[{{Oser} {et~al}\mbox{.}(2010){Oser}, {Ostriker}, {Naab}, {Johansson},
  \& {Burkert}}]{oser}
{Oser} L., {Ostriker} J.~P., {Naab} T., {Johansson} P.~H., {Burkert} A., 2010,
  \apj, 725, 2312

\bibitem[{{Pastorello} {et~al}\mbox{.}(2014){Pastorello}, {Forbes}, {Foster},
  {Brodie}, {Usher}, {Romanowsky}, {Strader}, \& {Arnold}}]{pasto}
{Pastorello} N., {Forbes} D.~A., {Foster} C., {Brodie} J.~P., {Usher} C.,
  {Romanowsky} A.~J., {Strader} J., {Arnold} J.~A., 2014, \mnras, 442, 1003

\bibitem[{{Quilis} \& {Trujillo}(2013)}]{Quilis}
{Quilis} V., {Trujillo} I., 2013, \apjl, 773, L8

\bibitem[{{Ruiz} {et~al}\mbox{.}(2014){Ruiz}, {Trujillo}, \&
  {M{\'a}rmol-Queralt{\'o}}}]{ruiz}
{Ruiz} P., {Trujillo} I., {M{\'a}rmol-Queralt{\'o}} E., 2014, \mnras, 442, 347

\bibitem[{{S{\'a}nchez} {et~al}\mbox{.}(2012){S{\'a}nchez}, {Kennicutt}, {Gil
  de Paz}, {van de Ven}, {V{\'{\i}}lchez}, {Wisotzki}, {Walcher}, {Mast},
  {Aguerri}, {Albiol-P{\'e}rez}, {Alonso-Herrero}, {Alves}, {Bakos},
  {Bart{\'a}kov{\'a}}, {Bland-Hawthorn}, {Boselli}, {Bomans},
  {Castillo-Morales}, {Cortijo-Ferrero}, {de Lorenzo-C{\'a}ceres}, {Del Olmo},
  {Dettmar}, {D{\'{\i}}az}, {Ellis}, {Falc{\'o}n-Barroso}, {Flores},
  {Gallazzi}, {Garc{\'{\i}}a-Lorenzo}, {Gonz{\'a}lez Delgado}, {Gruel},
  {Haines}, {Hao}, {Husemann}, {Igl{\'e}sias-P{\'a}ramo}, {Jahnke}, {Johnson},
  {Jungwiert}, {Kalinova}, {Kehrig}, {Kupko}, {L{\'o}pez-S{\'a}nchez},
  {Lyubenova}, {Marino}, {M{\'a}rmol-Queralt{\'o}}, {M{\'a}rquez}, {Masegosa},
  {Meidt}, {Mendez-Abreu}, {Monreal-Ibero}, {Montijo}, {Mour{\~a}o},
  {Palacios-Navarro}, {Papaderos}, {Pasquali}, {Peletier}, {P{\'e}rez},
  {P{\'e}rez}, {Quirrenbach}, {Rela{\~n}o}, {Rosales-Ortega}, {Roth},
  {Ruiz-Lara}, {S{\'a}nchez-Bl{\'a}zquez}, {Sengupta}, {Singh}, {Stanishev},
  {Trager}, {Vazdekis}, {Viironen}, {Wild}, {Zibetti}, \& {Ziegler}}]{califa}
{S{\'a}nchez} S.~F. {et~al.}, 2012, \aap, 538, A8

\bibitem[{{S{\'a}nchez-Bl{\'a}zquez}
  {et~al}\mbox{.}(2007){S{\'a}nchez-Bl{\'a}zquez}, {Forbes}, {Strader},
  {Brodie}, \& {Proctor}}]{pat}
{S{\'a}nchez-Bl{\'a}zquez} P., {Forbes} D.~A., {Strader} J., {Brodie} J.,
  {Proctor} R., 2007, \mnras, 377, 759

\bibitem[{{Schiavon} {et~al}\mbox{.}(2000){Schiavon}, {Barbuy}, \& {Bruzual
  A.}}]{Schiavon:00}
{Schiavon} R.~P., {Barbuy} B., {Bruzual A.} G., 2000, \apj, 532, 453

\bibitem[{{Smith}(2014)}]{smith}
{Smith} R.~J., 2014, \mnras, 443, L69

\bibitem[{{Spiniello} {et~al}\mbox{.}(2014){Spiniello}, {Trager}, {Koopmans},
  \& {Conroy}}]{Spiniello2013}
{Spiniello} C., {Trager} S., {Koopmans} L.~V.~E., {Conroy} C., 2014, \mnras,
  438, 1483

\bibitem[{{Spiniello} {et~al}\mbox{.}(2015){Spiniello}, {Trager}, \&
  {Koopmans}}]{Spiniello2014}
{Spiniello} C., {Trager} S.~C., {Koopmans} L.~V.~E., 2015, \apj, 803, 87

\bibitem[{{Stringer} {et~al}\mbox{.}(2015){Stringer}, {Trujillo}, {Dalla
  Vecchia}, \& {Martinez-Valpuesta}}]{Martin}
{Stringer} M., {Trujillo} I., {Dalla Vecchia} C., {Martinez-Valpuesta} I.,
  2015, \mnras, 449, 2396

\bibitem[{{Treu} {et~al}\mbox{.}(2010){Treu}, {Auger}, {Koopmans}, {Gavazzi},
  {Marshall}, \& {Bolton}}]{treu}
{Treu} T., {Auger} M.~W., {Koopmans} L.~V.~E., {Gavazzi} R., {Marshall} P.~J.,
  {Bolton} A.~S., 2010, \apj, 709, 1195

\bibitem[{{Trujillo} {et~al}\mbox{.}(2009){Trujillo}, {Cenarro}, {de
  Lorenzo-C{\'a}ceres}, {Vazdekis}, {de la Rosa}, \& {Cava}}]{truji09}
{Trujillo} I., {Cenarro} A.~J., {de Lorenzo-C{\'a}ceres} A., {Vazdekis} A., {de
  la Rosa} I.~G., {Cava} A., 2009, \apjl, 692, L118

\bibitem[{{Trujillo} {et~al}\mbox{.}(2014){Trujillo}, {Ferr{\'e}-Mateu},
  {Balcells}, {Vazdekis}, \& {S{\'a}nchez-Bl{\'a}zquez}}]{truji14}
{Trujillo} I., {Ferr{\'e}-Mateu} A., {Balcells} M., {Vazdekis} A.,
  {S{\'a}nchez-Bl{\'a}zquez} P., 2014, \apjl, 780, L20

\bibitem[{{Trujillo} {et~al}\mbox{.}(2011){Trujillo}, {Ferreras}, \& {de La
  Rosa}}]{truji11}
{Trujillo} I., {Ferreras} I., {de La Rosa} I.~G., 2011, \mnras, 415, 3903

\bibitem[{{van den Bosch} {et~al}\mbox{.}(2012){van den Bosch}, {Gebhardt},
  {G{\"u}ltekin}, {van de Ven}, {van der Wel}, \& {Walsh}}]{remco}
{van den Bosch} R.~C.~E., {Gebhardt} K., {G{\"u}ltekin} K., {van de Ven} G.,
  {van der Wel} A., {Walsh} J.~L., 2012, \nat, 491, 729

\bibitem[{{van Dokkum} \& {Conroy}(2010)}]{vandokkum}
{van Dokkum} P.~G., {Conroy} C., 2010, \nat, 468, 940

\bibitem[{{van Dokkum} {et~al}\mbox{.}(2010){van Dokkum}, {Whitaker},
  {Brammer}, {Franx}, {Kriek}, {Labb{\'e}}, {Marchesini}, {Quadri}, {Bezanson},
  {Illingworth}, {Muzzin}, {Rudnick}, {Tal}, \& {Wake}}]{vdk}
{van Dokkum} P.~G. {et~al.}, 2010, \apj, 709, 1018

\bibitem[{{Vazdekis} {et~al}\mbox{.}(1996){Vazdekis}, {Casuso}, {Peletier}, \&
  {Beckman}}]{vazdekis96}
{Vazdekis} A., {Casuso} E., {Peletier} R.~F., {Beckman} J.~E., 1996, \apjs,
  106, 307

\bibitem[{{Vazdekis} {et~al}\mbox{.}(2015){Vazdekis}, {Coelho}, {Cassisi},
  {Ricciardelli}, {Falc{\'o}n-Barroso}, {S{\'a}nchez-Bl{\'a}zquez}, {La
  Barbera}, {Beasley}, \& {Pietrinferni}}]{alpha}
{Vazdekis} A. {et~al.}, 2015, \mnras, 449, 1177

\bibitem[{{Vazdekis} {et~al}\mbox{.}(1997){Vazdekis}, {Peletier}, {Beckman}, \&
  {Casuso}}]{vazdekis:97}
{Vazdekis} A., {Peletier} R.~F., {Beckman} J.~E., {Casuso} E., 1997, \apjs,
  111, 203

\bibitem[{{Vazdekis} {et~al}\mbox{.}(2012){Vazdekis}, {Ricciardelli},
  {Cenarro}, {Rivero-Gonz{\'a}lez}, {D{\'{\i}}az-Garc{\'{\i}}a}, \&
  {Falc{\'o}n-Barroso}}]{miuscat}
{Vazdekis} A., {Ricciardelli} E., {Cenarro} A.~J., {Rivero-Gonz{\'a}lez} J.~G.,
  {D{\'{\i}}az-Garc{\'{\i}}a} L.~A., {Falc{\'o}n-Barroso} J., 2012, \mnras,
  424, 157

\bibitem[{{Vazdekis} {et~al}\mbox{.}(2010){Vazdekis},
  {S{\'a}nchez-Bl{\'a}zquez}, {Falc{\'o}n-Barroso}, {Cenarro}, {Beasley},
  {Cardiel}, {Gorgas}, \& {Peletier}}]{miles}
{Vazdekis} A., {S{\'a}nchez-Bl{\'a}zquez} P., {Falc{\'o}n-Barroso} J.,
  {Cenarro} A.~J., {Beasley} M.~A., {Cardiel} N., {Gorgas} J., {Peletier}
  R.~F., 2010, \mnras, 404, 1639

\bibitem[{{Weidner} {et~al}\mbox{.}(2013){Weidner}, {Ferreras}, {Vazdekis}, \&
  {La Barbera}}]{weidner:13}
{Weidner} C., {Ferreras} I., {Vazdekis} A., {La Barbera} F., 2013, \mnras, 435,
  2274

\end{thebibliography}

\appendix

\begin{center}
\section{Robustness of the inferred IMF profile}
\end{center}

To check the reliability of the IMF gradient of NGC\,1277 (Fig.~\ref{fig:imf}), we have explored three possible sources of systematics, namely, errors on telluric absorption correction, the effect of non-solar elemental abundances, and the nebular emission contamination to the H$_{\beta_\mathrm{o}}$ line. These tests -- supporting the robustness of our results -- have been performed because of some discrepancies between the best-fitting and observed line-strengths (see Sec.~\ref{sec:fitting}), in particular regarding the H$_{\beta_\mathrm{o}}$, Ca and Na spectral lines (see panels (c), (i) and (j) of Fig.~\ref{fig:ind}). 

\begin{figure}
\begin{center}
\includegraphics[width=8.cm]{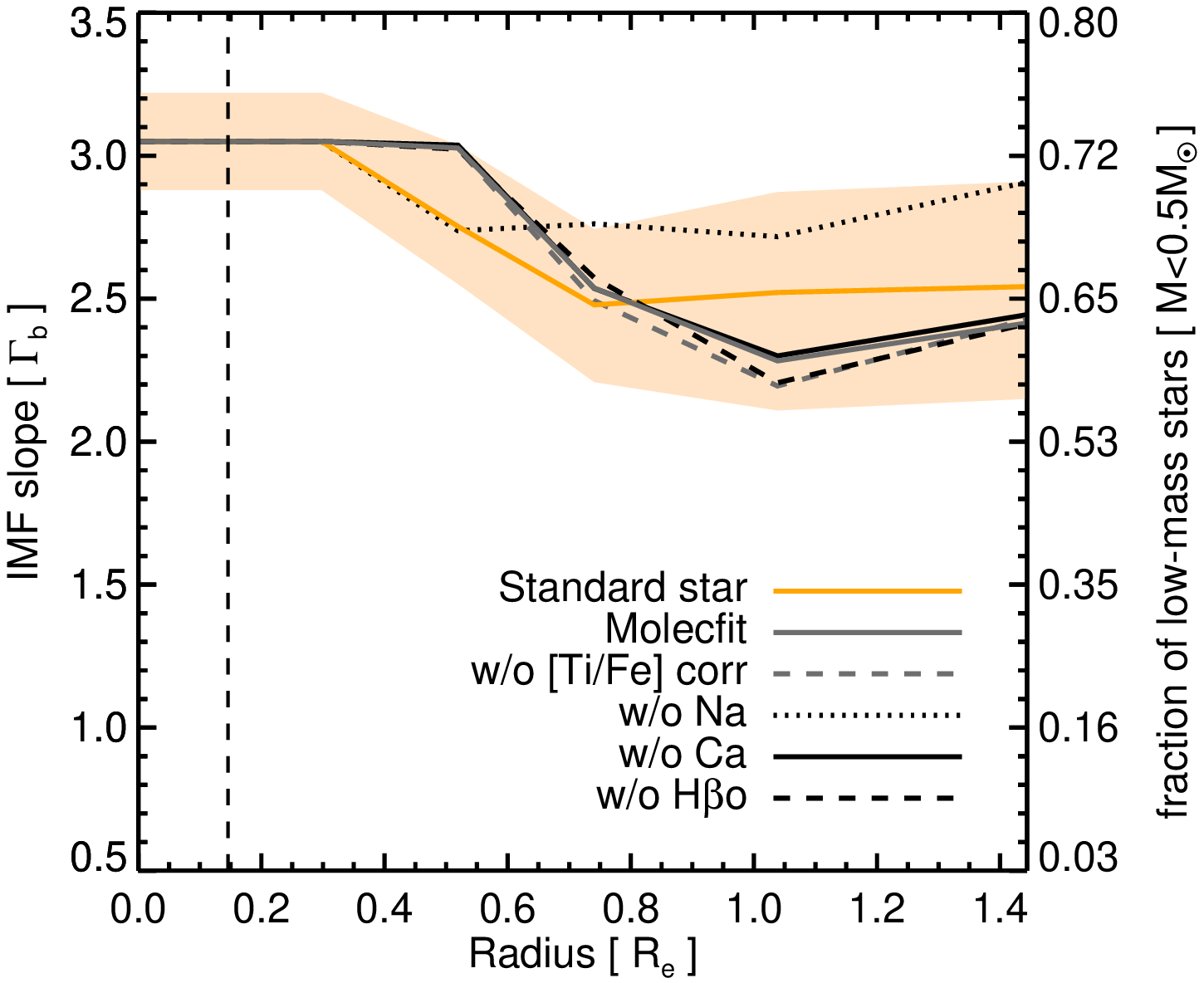}
\end{center}
\caption{Same as Fig.~\ref{fig:imf} but including tests to check the robustness of our IMF inference. The gray solid line indicates the IMF profile obtained using Molecfit to correct telluric absorption, whereas the gray dashed line corresponds to the best-fitting solution if the [Ti/Fe] dependence of H$_{\beta_\mathrm{o}}$ ignored. In solid black, doted and dashed lines we show the IMF profile when excluding from the fitting the Calcium, Sodium and H$_{\beta_\mathrm{o}}$ features, respectively. The overall agreement among all the fits shows that our main result is not significantly affected by telluric, nebular emission or model uncertainties.  Dashed vertical lines mark the seeing radius.}
\label{fig:tests}
\end{figure}

\begin{figure}
\begin{center}
\includegraphics[width=8.cm]{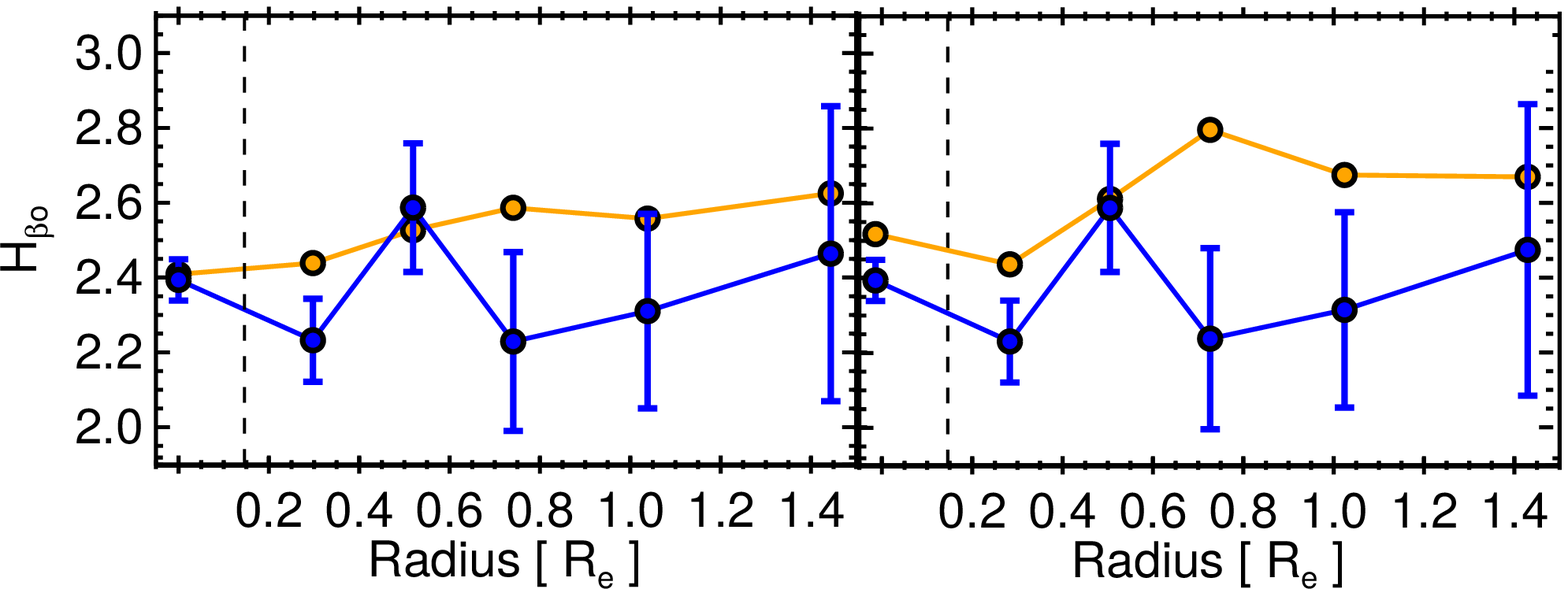}
\end{center}
\caption{Observed and best-fitting H$_{\beta_\mathrm{o}}$ radial profiles. The right panel is the same as panel (c) in Fig.~\ref{fig:ind}, whereas the left panel corresponds to the case where [Ti/Fe] abundance effects on H$_{\beta_\mathrm{o}}$ are neglected. Notice, in the latter case, the better agreement between best-fitting and observed indices, with no significant impact on the inferred IMF profile (Fig.~\ref{fig:tests}).}
\label{fig:hbo}
\end{figure}

The fact that the best-fitting H$_{\beta_\mathrm{o}}$ lies above the observed values (panel (c) in Fig.~\ref{fig:ind}) may indicate some residual contamination from nebular emission to the line profile. In the present work, the nebular emission correction to H$_{\beta_\mathrm{o}}$ has been estimated in a self-consistent manner (see La Barbera et al.~2013 for details), i.e., for each IMF slope ($\Gamma_\mathrm{b}$) explored in the fitting process, we apply the correction calculated by simultaneously fitting the H$_{\beta}$ line profile with a 2 SSP model (for a given $\Gamma_\mathrm{b}$) plus a gaussian emission line.  For the two innermost radial bins, 
the correction turned out to be fully consistent with zero regardless of 
$\Gamma_\mathrm{b}$, while in the bins from $\sim0.5$ to $1.5$~\re, a mild correction
was found~\footnote{Notice that this correction is already applied to observed H$_{\beta_\mathrm{o}}$ 
values (blue dots) in Fig.~\ref{fig:ind}. }, in the range from $0.1$ to $0.2 \, \AA$. Therefore, it is unlikely that differences in panel (c) of Fig.~\ref{fig:ind} are due to residual nebular contamination. Note also that underestimating H$_{\beta_\mathrm{o}}$ (in particular for the outermost bins) would imply that, given the age sensitivity of the TiO features, the inferred IMF slope is shallower than the ``true'' one, hence strengthening our conclusion about a shallow IMF gradient in NGC\,1277.
The possible explanation for the mismatch between the best-fitting and observed H$_{\beta_\mathrm{o}}$  may reside in the uncertainty of the stellar population models used to account for  non-solar abundance ratios on line strengths (in particular H$_{\beta_\mathrm{o}}$). We modelled deviations from solar partition with the semi-empirical approach described in \citet{labarbera}, with the further aid of the \citet[hereafter CvD]{cvd12} stellar population models. In practice, the $\Delta_{\alpha,i}$ terms in Eq.~\ref{eq:method} can be expanded as follows \citep[see \S3.3 in][]{mn14}:

\begin{equation*}
 \Delta_{\alpha,i} = C_{\alpha,i} \cdot [\alpha/{\rm Fe}] + \Delta_{\rm X,i} \cdot \delta {\rm X}
\end{equation*}

\noindent where $C_{\alpha,i}$ is the observed sensitivity of the $i$-th index to [$\alpha$/Fe] (derived from SDSS stacked spectra, as in \citealt{labarbera}); $\Delta_{\rm X,i}$ is the theoretical sensitivity of each index to the abundance ratio of a specific element X, 
estimated with CvD models, and $\delta {\rm X}$ is a ``residual''  abundance of element X, not accounted for by the $C_{\alpha,i}$ correction, and treated as a fitting parameter. In other terms, we remove the bulk dependence of each index on [$\alpha$/Fe] with the $C_{\alpha,i}$ terms, while still allowing for some residual abundance effect on the indices through the best-fitting $\delta {\rm X}$'s. In practice,  following \citealt{labarbera}, we have considered only the effect of elemental abundances of selected gravity-sensitive features, i.e. $\rm X=Ca$, Na, and Ti. Regarding H$_{\beta_\mathrm{o}}$, we have $\Delta {\rm X} \sim 0$ for all X's but $\rm X=Ti$, i.e. CvD models predict a significant sensitivity of H$_{\beta_\mathrm{o}}$ to [Ti/Fe] abundance ratio (in the sense of  H$_{\beta_\mathrm{o}}$ increasing with [Ti/Fe]). In fact, we found that the [Ti/Fe] tends, on average, to increase the  best-fitting H$_{\beta_\mathrm{o}}$ by $\sim$0.15\AA, possibly explaining at least part of the observed mismatch between the best-fitting~\footnote{Note that the best-fitting indices in Fig.~\ref{fig:ind} are actually given by $\rm EW_{M,i}+\Delta_{\alpha,i}$.} and observed H$_{\beta_\mathrm{o}}$ in panel (c) of Fig.~\ref{fig:ind}. Since the dependence of H$_{\beta_\mathrm{o}}$ on specific elemental abundances may be significantly affected by theoretical uncertainties on stellar population models \citep[see][]{Cervantes},
we have repeated the fitting process by (i) removing the $\Delta_{\rm Ti,\mathrm{H}_{\beta_\mathrm{o}}}\cdot \delta {\rm Ti}$ term from Eq.~\ref{eq:method}, and (ii)
excluding the H$_{\beta_\mathrm{o}}$ index altogether from the fits. The resulting IMF profiles from these tests are plotted in Fig.~\ref{fig:tests}, showing no significant differences with respect to the reference profile in Fig.~\ref{fig:imf}. Moreover, neglecting the [Ti/Fe] effect on H$_{\beta_\mathrm{o}}$ leads to a better agreement between observed and best-fitting radial trends for this line, as shown in Fig.~\ref{fig:hbo}.

As mentioned in \S~\ref{sec:data}, one major issue in constraining the IMF is the removal of telluric 
lines from the observed spectra. In the present work, we have removed atmospheric absorption with two
independent approaches, i.e., (1) using a spectrophotometric standard star and (2) applying the software  {\sc MOLECFIT}. Fig.~\ref{fig:tests} compares the corresponding IMF best-fitting profiles,
showing no significant differences within the uncertainties. The two approaches lead to 
an estimated IMF radial variation of (1) $\Delta\Gamma_\mathrm{b}= -0.5$ and (2) $\Delta\Gamma_\mathrm{b}= -0.6$, respectively. 

To further explore to what extent the mismatch of Ca and Na lines for some radial bins (see panels (i) and (j) of Fig.~3) may be affecting our conclusions, we have recomputed the IMF profile by excluding, as we did for H$_{\beta_\mathrm{o}}$ (see above), all Na and Ca lines in turn, from the fitting procedure. Again, the resulting IMF gradients, shown in Fig.~\ref{fig:tests}, are compatible with the IMF profile reported in Fig.~\ref{fig:imf}. It is worth noting that the impact of removing Na lines is the most significant, suggesting either some residual telluric contamination in the data, or some model uncertainties in dealing with the abundance ratios \citep{Spiniello2014}. 

All the above tests, summarized in Fig.~\ref{fig:tests}, indicate that our IMF inference is robust against \emph{i)} nebular emission correction,\emph{ii)}, abundance effects (on H$_{\beta}$ line), \emph{iii)} age determination through H$_{\beta_\mathrm{o}}$ itself, \emph{iv)} telluric correction, and \emph{v)} the set of lines included in the analysis. 

\label{lastpage}

\end{document}